\documentclass{elsart}
 \usepackage{graphicx,cite}

\begin{document}
 \begin{frontmatter}
 \title{A density matrix approach to photoinduced electron
  injection}
  \author{Michael Schreiber, Ivan Kondov,  Ulrich  Kleinekath\"ofer\thanksref{ca}} 
 \thanks[ca]{Fax: +49 371 531 3151, e-mail:
  kleinekathoefer@physik.tu-chemnitz.de}

 \address{Institut f\"ur Physik, Technische Universit\"at, 
 D-09107 Chemnitz, Germany}

 \journal{Journal of Luminescence}
 \date{\today}

 \begin{abstract}
   Electron injection from an adsorbed molecule to the substrate
   (heterogeneous electron transfer) is studied. One reaction coordinate is
   used to model this process. The surface phonons and/or the electron-hole
   pairs together with the internal degrees of freedom of the adsorbed
   molecule as well as possibly a liquid surrounding the molecule provide a
   dissipative environment, which may lead to dephasing, relaxation, and
   sometimes excitation of the relevant system. In the process studied the
   adsorbed molecule is excited by a light pulse. This is followed by an
   electron transfer from the excited donor state to the quasi-continuum of
   the substrate. It is assumed that the substrate is a semiconductor. The
   effects of dissipation on electron injection are investigated.

 \end{abstract}


 \end{frontmatter}

 \section{Introduction}

 In recent years electron transfer (ET) between molecular adsorbates and
 semiconductor nanomaterials and surfaces has been subject of much research
 \cite{asbu01}. The injection of an electron into the conduction band is
 a prototype reaction for a lot of electrochemical and photoelectrochemical
 interfacial processes such as photography, solar energy conversion, 
 quantum dot devices, etc.\  \cite{asbu01}.
 Interfacial ET between discrete molecular levels and a conducting surface
 is the simplest of all surface reactions: it involves only the exchange of
 an electron, and so no bonds are broken \cite{lanz94}. 

 The ultrafast nature of the charge injection from adsorbed molecules to
 the conduction band of semiconductor surfaces was shown in recent experiments
 \cite{burf96,cher97,hann97}. The theoretical description of such
 experiments demands an adequate treatment of the ET dynamics to be able to
 describe short time-scale phenomena such as coherences. This can be done within the reduced
 density matrix (RDM) description used in the present contribution.

 Recently \cite{rama00b,rama00} the electron injection from a chromophore to
 a semiconductor conduction band was described using the time-dependent
 Schr\"odinger equation, thus neglecting relaxation processes. The neglect of
 relaxation processes was motivated by the experimental finding that
 injected electrons relax only within 150 fs in the perylene-TiO$_2$
 system. Here we include relaxation to be able to treat a larger class of
 experiments where, for example, the adsorbed molecule is surrounded by a
 liquid environment, and longer times.

 \section{Theory}

 In the RDM theory the full system is divided into a relevant system part
 and a heat bath. Therefore the total Hamiltonian consists of three terms
 -- the system part $H_{\rm S}$, the bath part $H_{\rm B}$, and the
 system-bath interaction $H_{\rm SB}$:
 \begin{equation}
 H = H_{\rm S} + H_{\rm B} + H_{\rm SB}.
 \label{eq:Hamiltonian}
 \end{equation}
 The RDM $\rho$ is obtained from the density matrix of the full system by
 tracing out the degrees of freedom of the environment.  This reduction
 together with a second-order perturbative treatment of $H_{\rm SB}$ and the
 Markov approximation leads to the Redfield equation
 \cite{redf57,redf65,blum96,may00}:
 \begin{equation}
   \dot{\rho} = - i [H_{\rm S},\rho] + {\mathcal R} \rho ={\mathcal L} \rho.
 \label{eq:redfield}
 \end{equation}
 In this equation ${\mathcal R}$ denotes the Redfield tensor. If one assumes
 bilinear system-bath coupling with system part $K$ and bath part $\Phi$
 \begin{equation}
 H_{\rm SB} = K \Phi
 \label{eq:bath-coupling}
 \end{equation}
 one can take advantage of the following decomposition \cite{poll94,may00}:
 \begin{equation}
 \dot{\rho} = - i \left[ H_{\rm S},\rho \right]
 + 
 [ \Lambda\rho,K]+
 [ K,\rho\Lambda^{\dagger}]
 .
 \label{eq:pf-form}
 \end{equation}
 The $\Lambda$ operator can be  written in the form
 \begin{equation}
 \Lambda=\int\limits_0^{\infty}d\tau
 \langle\Phi(\tau)\Phi(0)\rangle K^{\rm{I}}(-\tau)
 \label{eq:lambda}
 \end{equation}
 where $ K^{\rm{I}}(-\tau)=  e^{-i H t} K e^{i H t}$ is
 the operator $K$ in the interaction representation.

 The system bath interaction is taken to be linear in the reaction
 coordinate as well as in the bath coordinates. Neither the rotating wave
 nor the secular approximation have been invoked. The so-called diabatic
 damping approximation which has numerical advantages \cite{kond00}
 is not used because it could lead to wrong results in the present system
 studied \cite{egor01,klei01a}. 

 In the following we direct our attention to ET 
 between an excited molecular state and a conduction band.
 The Hamiltonian modeling this system consists of the ground and one
 excited state of the molecule and a quasi-continuum describing the
 conduction band together with one vibrational coordinate
 \begin{equation}
   \label{ham}
   H=\sum_a H_a |\phi_a \rangle \langle \phi_a|
 + \sum_k (V_{ke}  |\phi_k \rangle \langle \phi_e| +H.c.)~.
 \end{equation}
 Here $a$ can be equal to $g$ for the ground state, $e$ for the excited
 state, and $k$ for the quasi-continuum. As in Ref.\ \cite{rama00b} we
 choose the frequency of the vibrational mode to be $\hbar \omega_{\rm
   vib}=0.1~{\rm eV}$.  The coupling between the excited state and the continuum
 states is assumed to be constant: $V_{ek}=0.1~{\rm eV}$. A box-shaped uniform
 density of states is used.  Instead of modeling the excitation from the
 ground state explicitly we assume a $\delta$-pulse. The excited
 state potential energy surface is shifted 0.1~\AA{} along the reaction
 coordinate with respect to the ground state potential energy surface. 
 This results in an initial vibrational wave packet on the excited state
 with significant population in the lowest 4~-~5 vibrational states.  The
 shift between the excited state energy surface and the continuum parabola is
 0.2~\AA{}.  The thermal bath is characterized by its spectral density
 $J(\omega) = \sum_m \gamma_m \delta(\omega-\omega_m)$.  Because all system
 oscillators have the same frequency the coupling to the bath can be given
 by one parameter $\gamma_1$ in the diabatic damping approximation.
 Denoting the effective mass of the harmonic oscillator by ${\mathcal M}$
 the strength of the damping is chosen as
 $\gamma_1 \pi/({\mathcal M} \omega_{\rm vib})=0.1~{\rm eV}$.

 To be able to study the effects of dissipation we do not model the
 quasi-continuum with such a large number of electronic states as in
 Ref.\ \cite{rama00b}. In that work a band of width 2 eV was described
 using an energy difference of 2.5 meV leading to 801 electronic surfaces.
 These calculations are already demanding using wave packet propagation but
 almost impossible using direct density matrix propagation. For doing such
 a large system one would have to use the Monte Carlo wave function scheme
 \cite{wolf95,wolf96}.  We use a much simpler model and describe only
 that part of the conduction band which really takes part in the injection
 process. The total width of the conduction band may be significantly
 larger. In the following, a band of
 width 0.75 eV is treated with 31 electronic surfaces. In each of these
 electronic states five vibrational states are taken into account.
We are aware that this is only a minimal model but hope that it catches the
 effects of dissipation on the electron injection process.

 \section{Results}

 Here we look at two different populations arising in the process of
 electron injection. The time-dependent
 population of the electronic states in the conduction band is calculated as the sum over the
 vibrational levels of each electronic surface $P(k,t)=\sum_{\nu}
 P_{k,\nu}(t)$.
 As a second quantity we look at the time-dependent
 population of the vibrational levels of the excited molecular state
 $P_e(\nu,t)$.
 These two probability distributions give some hints on the effect of
 dissipation. 

 Figure\ 1 shows the electronic population for the quasi-continuum, i.e.\ the
 probability distribution of the injected electron, versus the energy of the
 conduction band. As described above, the four lowest vibrational states are
 populated significantly at $t=0$. The structure arising in the upper panel
 of Fig.~1 was already explained by Ramakrishna et al.\ \cite{rama00b}.
 It can be estimated using the golden rule. The electronic probabilities in
 the quasi-continuum are given as 
 \begin{equation}
 P(k,t) \approx \sum_{\mu{},\nu} P_\mu{}^{(i)}
 |\langle \chi_{l \mu{}}| \chi_{k \nu} \rangle|^2
 \delta(E_0 + \mu{}\hbar \omega_{\rm vib} -E -\nu  \hbar \omega_{\rm vib})
 \end{equation}
 where $ P_\mu{}^{(i)}$ is the initial vibronic distribution in the excited
 state and $|\chi_{l \mu{}} \rangle$ and $|\chi_{k \nu} \rangle$ are the
 vibronic parts of the wave packet in the excited and quasi-continuum
 states, respectively. The energy $E_0$ denotes the middle of the band.
 Turning on dissipation two effects can be seen. First, the vibrational
 populations in the excited state of the molecule no longer only decay into
 the quasi-continuum states but also relax within the excited state (see
 Fig.~2). Second, the vibrational populations also relax within the
 quasi-continuum states. The recurrences back into the excited state become
 much smaller. Only those parts of the wave packet which are still high
 enough in energy can go back to the molecule.

 In summary, we extended the work by Ramakrishna, Willig, and May
 \cite{rama00b} by including relaxation processes into the description of
 electron injection into the conduction band of a semiconductor. This will,
 at least, 
 become important for modeling electron injection in the presence of a
 fluid surrounding the attached molecule.


 \begin{ack}
 Financial support of the DFG is gratefully acknowledged.
 \end{ack}




\newpage

\begin{figure}
\caption{Probability distribution of the injected
  electron $P(k,t)$ without dissipation (upper panel) and with
  dissipation (lower panel).}
\end{figure}
\begin{center}
\includegraphics[]{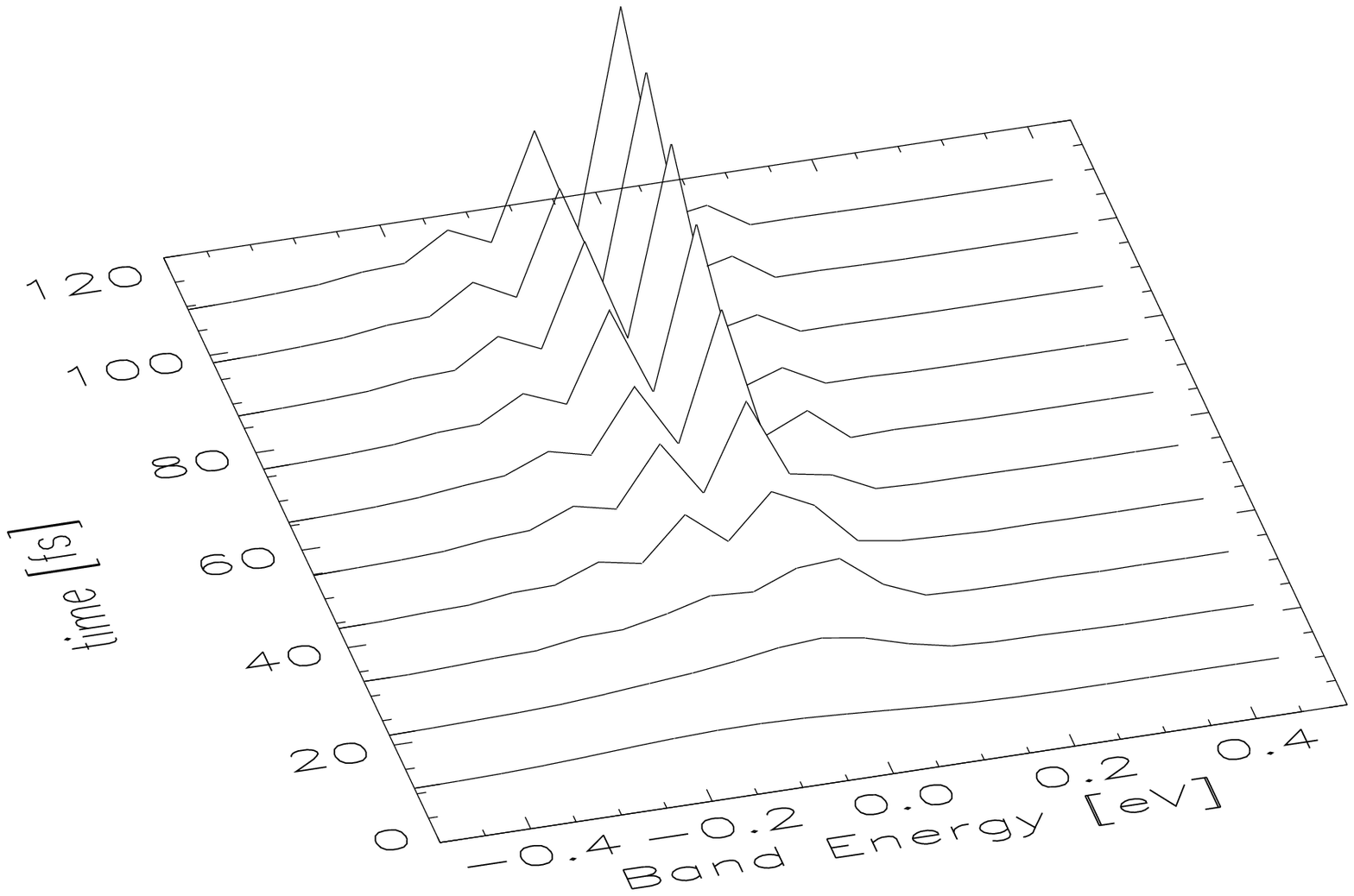}
\includegraphics[]{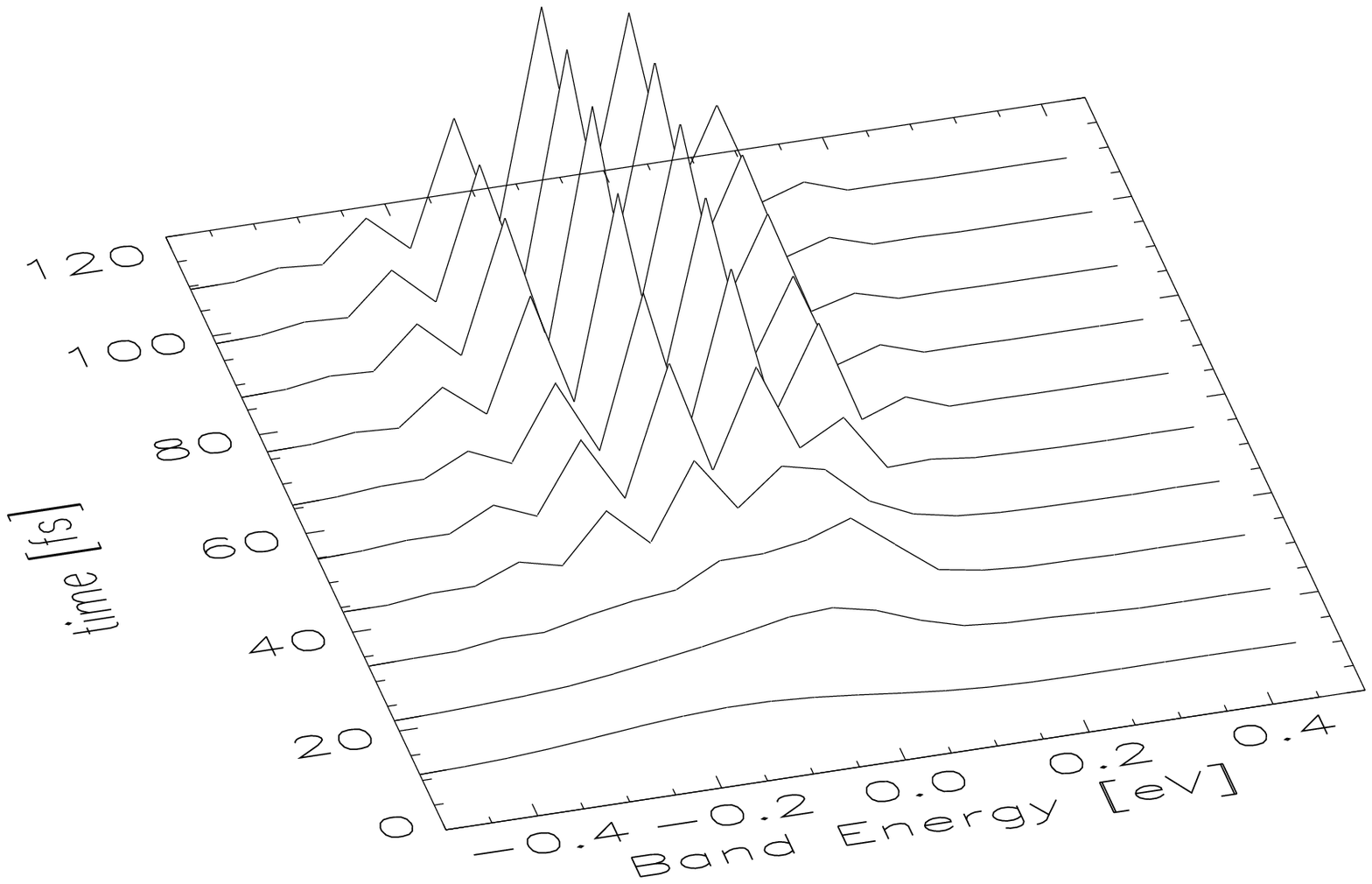}
\end{center}

\begin{figure}
\caption{Population of the vibrational levels of the excited molecular
  state $P_e(\nu,t)$ without dissipation (upper panel)  and with
  dissipation (lower panel).} The lowest vibrational state is populated
  most at $t=0$. The higher the vibrational quantum number the less
  populated is the level.
\end{figure}

\begin{center}
\includegraphics[width=12cm]{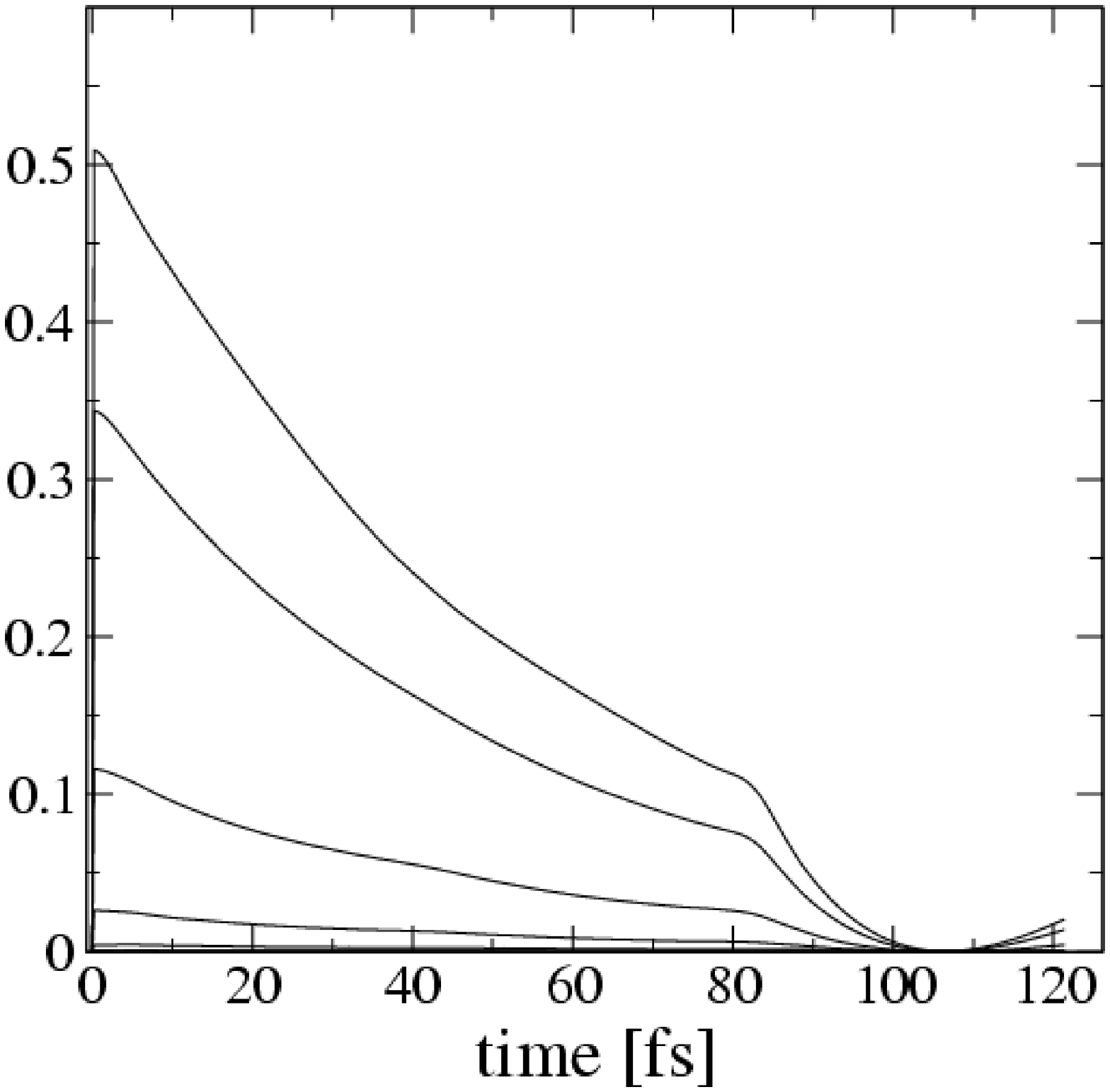}
\includegraphics[width=12cm]{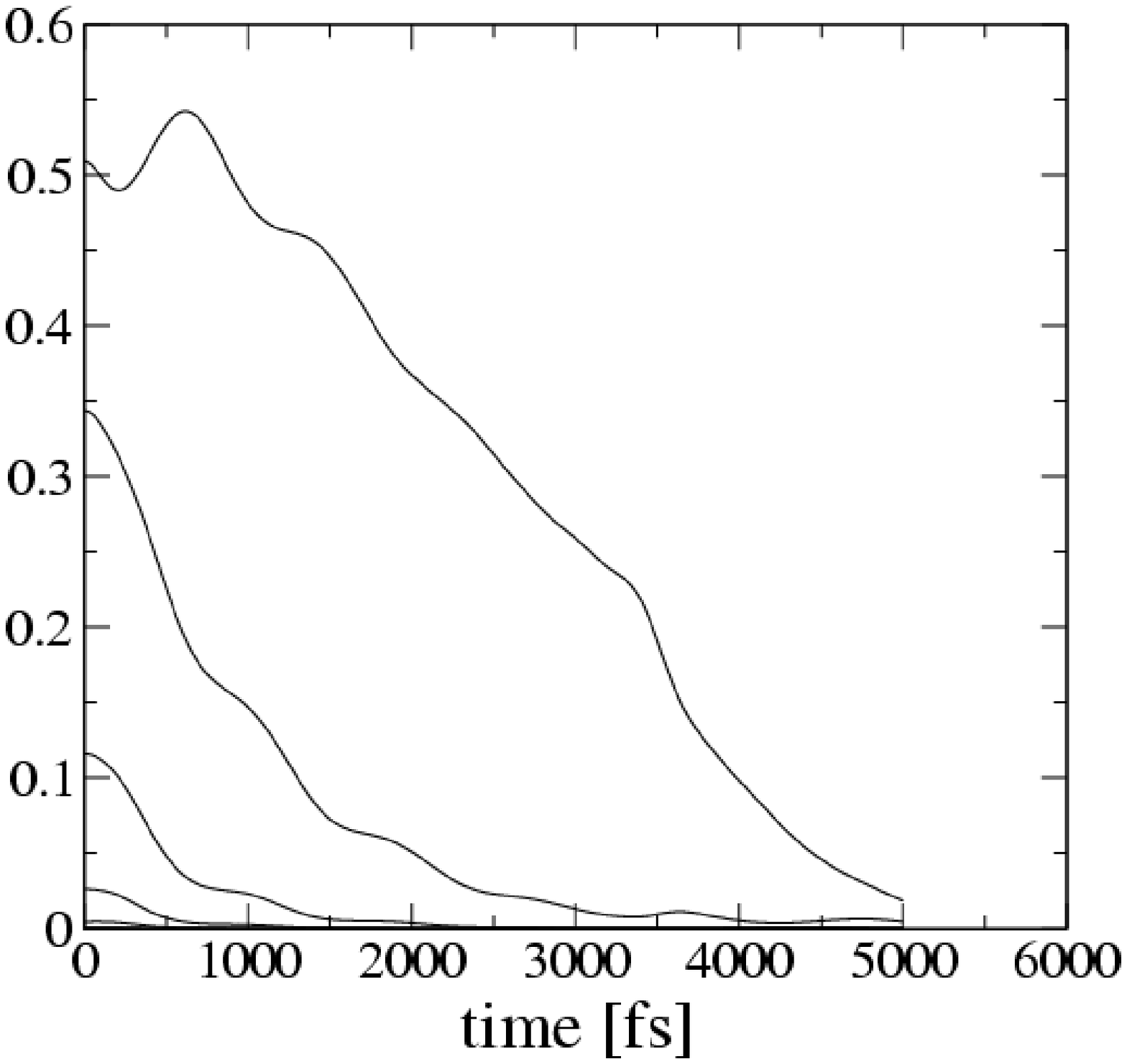}
\end{center}

\end{document}